\documentclass[final]{amsart}

\RequirePackage[utf8]{inputenc}
\RequirePackage[leqno]{mathtools}
\RequirePackage{amsthm}
\RequirePackage{amssymb}
\RequirePackage[mathcal]{euler}
\RequirePackage{dsfont}
\RequirePackage{color}

\newtheorem{thm}{Theorem}

\newcommand{\tsum}{\mathop{\textstyle \sum }}

\def\R{\mathds{R} }
\def\Q{\mathds{Q} }
\def\tendsto{\longrightarrow}

\def\defem#1{{\sl#1}}
\makeatletter
\@ifpackageloaded{hyperref}
{
	\def\mref#1#2{\hyperref[#2]{#1~\ref*{#2}}}
	\def\meqref#1#2{\hyperref[#2]{#1~(\ref*{#2})}}
	
} {
	\def\mref#1#2{#1~\ref{#2}}
	\def\meqref#1#2{#1~\eqref{#2}}
	
	\newcommand{\hyperref}[2][1]{#2}
}
\makeatother

\begin{document}

	\begin{abstract}
	We present a somewhat different way of looking on Shannon entropy. This leads to an axiomatisation
	of Shannon entropy that is essentially equivalent to that of Fadeev.\end{abstract}

\title{A note on Shannon entropy}
\author[T. Sobieszek]{Tomasz Sobieszek}
\email{sobieszek@math.uni.lodz.pl}
\urladdr{http://sobieszek.co.cc}
\address{
		Faculty of Mathematics and Computer Science\\
		University of Łódź\\
		ul. Banacha 22, 90-238 Łódź \\
		Poland
}

	\thanks{This paper is partially supported by Grant nr N N201 605840.}

\maketitle

	\section {Introduction}

{
{A large part of the discrete theory of information concerns itself with real functions $H$ defined on the family of sequences
$(p_1,\ldots,p_n)$ such that $p_i \ge 0$ and $\tsum p_i = 1$. There, a very significant role is played
by the \defem{Shannon entropy}, given by
\[
	H(p_1,\ldots,p_n)= p_1\log\tfrac 1{p_1} + \ldots + p_n \log \tfrac 1{p_n}.\footnotemark
\]
\footnotetext{In here, as throughout the
paper, we confine ourselves to base~$2$ logarithms, as dictated by information theory tradition.}
It is the only symmetric (i.e. independent of the order of $p_i$-s) continuous function
of such sequences that is normalised by $H(1/2,1/2)=1$ and satisfies the following \defem{grouping axiom}
\begin{multline}	\label{eqn:entropy}
	H(\ a_1 p_1, \ldots, a_k p_1\ , \  b_1 p_2, \ldots, b_l p_2\ , \ \cdots\ , \ c_1 p_n, \ldots, c_m p_n\ ) =
\\	
	\begin{aligned}
	&\ H (p_1, p_2, \cdots, p_n)\  +\\
	p_1 H(a_1, \ldots, a_k) &+ p_2 H(b_1,\ldots,b_l) + \cdots + p_n H(c_1,\ldots,c_m).
	\end{aligned}
\end{multline}
This result which is a better version of Shannon's own set of axioms (see~\cite{Sh}) is a slight modification
of Fadeev's axioms of entropy, c.f. \cite{Fa}.

The shape of the grouping axiom, leads us to think about entropy as a value assigned to transformations, divisions or partitions, say from a number $p_1$, to its partition $\ a_1 p_1, \ldots, a_k p_1$, where $a_i$ sum up to $1$. In fact, we will extend $H$ to nonnegative sequences, so that
$H(a_1 p_1 , \ldots, a_k p_1 ) = p_1 H(a_1, \ldots, a_k)$ and satisfies
\begin{multline*}	\label{eqn:entropy}
	H(\ a_1 p_1, \ldots, a_k p_1\ , \  b_1 p_2, \ldots, b_l p_2\ , \ \cdots\ , \ c_1 p_n, \ldots, c_m p_n\ ) =
\\	
	\begin{aligned}
	&\quad H (p_1, p_2, \cdots, p_n)\ + \\
	H(a_1 p_1 , \ldots, a_k p_1 ) &+ H(b_1 p_2,\ldots,b_l p_2) + \cdots + H(c_1 p_n,\ldots,c_m p_n).
	\end{aligned}
\end{multline*}
whenever $a_i$-s, $b_i$-s, $\ldots$, and $c_i$-s sum up to~$1$.

Our approach fits in with the  beautiful approach to entropy presented in~\cite{BaFrLe} a bit more naturally than the originally used Fadeev's system of axioms.

For a detailed exposition of Shannon entropy, related entropies and the various conditions related with
their definition, see~\cite{AD}. For a modern survey of characterisations of Shannon entropy (among other
things), see~\cite{Cs}.}}

	\section{Entropy as a homogenous quantity, additive on partitionings of positive numbers.}

Each function~$H$ on sequences $(p_1,\ldots,p_n)$, $p_i\ge 0$, $\sum p_i =1$ can be naturally extended to a homogenous\footnotemark \ function $\hat H$ on seqences $(a_1,\ldots,a_n)$, $a_i\ge 0$, $\sum a_i >0$ by setting
\[
	\hat H(a_1,\ldots,a_n)= sH\big(\tfrac{a_1}s,\ldots,\tfrac{a_n}s\big),\quad\text{where $s = a_1+\ldots +a_n$}.
\]
Conversely every function $\hat H$ on sequences of nonnegative reals can be restricted to the domain
of $H$. This lead to a bijective identification of $H$ and a homogenous $\hat H$.
\footnotetext{i.e. such that $\hat H(c a_1,\ldots, c a_n)  = c H(a_1,\ldots,a_n)$, for $c>0$.}

The function $H$  satisfies the entropy equation if and only if $\hat H$ satisfies the following
equation, closely related to \defem{2-cocycle equation} (c.f.~\cite{Eb}, and Remark 1 in~\cite{So};)
\begin{multline}	\label{eqn:cocycle}
	\hat H(\ a_1 , \ldots, a_k \ , \  b_1, \ldots, b_l \ , \ \cdots\ , \ c_1 , \ldots, c_m \ ) =
\\	
	\begin{aligned}
	&\ \hat H (a_1 +\ldots +a_k, b_1 + \ldots + b_l, \cdots, c_1 + \ldots + c_m) \\
	+ \ &\hat H(a_1, \ldots, a_k)+ \hat H(b_1,\ldots,b_l) + \cdots + \hat H(c_1,\ldots,c_m).
	\end{aligned}
\end{multline}

if  we interpret $\hat H(a_1, \ldots, a_k)$ as the `entropy' of the partitioning of
$a_1+ \ldots + a_k$ into $a_1,\ldots, a_k$, this equation expresses a kind of additivity, that
the entropy of the partitioning of
$a_1 +\ldots +a_k \ + \ b_1 + \ldots + b_l \ + \ \cdots \ + \ c_1 + \ldots + c_m$ into
$a_1 , \ldots, a_k \ , \  b_1, \ldots, b_l \ , \ \cdots\ , \ c_1 , \ldots, c_m$ is a sum of `entropies' of
the half-way partitionings that go through the groups $a_1+\ldots+a_k$, $b_1 + \ldots + b_l$, $\cdots$, $c_1 + \ldots + c_m$.

It is rather expected 
that a symmetric function $\hat H$ (not necessarily homogenous)
satisfies the 2-cocycle equation if and only if there exists a `potential'
function $g:[0,\infty)\to \R$ such that (see 
Lemma 1 in~\cite{So})
\begin{equation}\label{eqn:Hg}
	\hat H (a_1,\ldots, a_n) = g(a) + \ldots +g(a_n) \ - \ g(a_1 +\ldots + a_n).
\end{equation}
Moreover we can, and we will assume that $g(1) =0$. Since $H(1/2,1/2)=1$ we have $g(1/2)=1/2$.

Now, $\hat H$ is homogenous if and only if $\hat H(\cdot,\cdot)$ is homogenous. This in turn
is equivalent to
\begin{equation}	\label{eqn:ghomogenity}
	g(a(b_1+b_2)) -g(ab_1) -g(ab_2) = a \big[g(b_1+b_2) - g(b_1) - g(b_2)\big]
\end{equation}

Let $D$ be  a function on pairs of nonnegative numbers such that
\begin{equation}	\label{eqn:g}
	g(ab) = ag(b) + bg(a) + D(a,b).
\end{equation}
It follows that $g$ satisfies \meqref{equation} {eqn:ghomogenity} if and only if
$D(a,\cdot)=D(\cdot,a)$ is additive i.e. if and only if $D$ is $\Q_+$-bilinear.

Assume now that $\hat H$ is derived from $H$ i.e. that $\hat H$ is homogenous. Then $D$ is $\Q_+$-bilinear.
Since $g(1)=0$ we conclude that
\begin{equation}	\label{eqn:grational}
	g(ab) = ag(b) + bg(a)\quad\text{for nonnegative rational $a$, and $b$.}
\end{equation}

Now, let us remark that if we could prove that $g$ is continuous it would follow that
\meqref{equation} {eqn:grational} is satified for all nonnegative $a$, $b$. Then the
function $l$ such that $g(a) = a l(a)$ would be continuous and would satisfy the  equation
\begin{equation}\label{eqn:log}
	l(ab)=l(a)+ l(b),
\end{equation}
and therefore would be a logarithm, $l(x) = c\cdot \ln x$. Since $l(1/2)=1$, we would have
$l(x)=\log (1/x)$, and $g(x)=x\log (1/x)$ i.e. we would show that $H$ is a Shannon entropy.

Assume that $H$ is continuous, or equivalently that $\hat H$ is continuous. The proof that $g$
is continuous, or that $l$ is a logarithm is the crucial part of the reasoning. The author of this paper
cannot find an easier way, or for that matter any significantly different proof, than the one that makes
use of the following theorem (c.f.~\cite{Re}, Ch IX,Theorem 2, p. 544)

		\begin{thm}
	Let $l:\{1,2,\ldots\}\to \R$ be a function that satisfies the following conditions
	\begin{align}
		&l(ab)=l(a)+l(b)\label{eqn:lErdos}\\
		&l(n+1) - l(n) \tendsto 0,\quad \text{as $n$ tends to $\infty$.}\label{eqn:lErdos2}
	\end{align}
	Then $l(n) = c\cdot \ln n$. \end{thm}

By using this theorem our reasoning converges with the proof (as given in~\cite{Re}, Ch~IX) that the Fadeev axiomatisation
uniquely describes Shannon entropy. In fact, from the continuity of $\hat H$ we have
\begin{align*}
	l(n+1) - l(n) + \frac {l(n+1)}n &=\\
		&= g\big(\frac {n+1}n\big) -g(1) - g\big(\frac 1n\big) \tendsto 0,
\end{align*}
which is almost \eqref{eqn:lErdos2}. The only element missing is supplied by the following elementary
result due to Mercer (see~\cite{Ha}, or~\cite{Re})

		\begin{thm}
Let $a_n$ be a sequence of numbers, and let $s_n=a_1+\ldots + a_n$ be its partial sum. Then
\[
	a_n \tendsto a\quad\text{if and only if}\quad a_n + \frac {s_n}n \tendsto 2a	.
\]
\end{thm}
We just need to use it with $a_n=l(n+1)-l(n)$, and $a=0$. We infer that $l(a) = \log (1/a)$, and
$g(a) = a \log (1/a)$ for rational $a>0$. Denote $u(a):=a \log (1/a)$ for all $a>0$. Then from \meqref
{equation} {eqn:Hg} for rational $a_i$ and the continuity of $H$ and $u$ we infer that
\[
\hat H(a_1,\ldots, a_n)=u(a_1)+\ldots+u(a_n) - u(a_1 + \ldots + a_n).
\]

This concludes our reasoning. Gathering it all together, we have shown the following theorem, essentially equivalent to the Fadeev axiomatization of
entropy:

		\begin{thm}
	Let  $\hat H$ be a function that satisfies the following conditions:
	\begin{enumerate}
		\item $\hat H$ is homogenous,
		\item $\hat H$ is symmetric,
		\item $\hat H$ satisfies the 2-cocycle equation \eqref{eqn:cocycle},
		\item $\hat H$ is continuous,
		\item $\hat H(1,1) = 2$.
	\end{enumerate}
	Then $\hat H(a_1,\ldots, a_n)=u(a_1)+\ldots+u(a_n) - u(a_1 + \ldots + a_n)$, where
	$u(x) =x \log (1/x)$.
	\end{thm}


These notes are the expression of my unwritten view of Shannon entropy that I held for a few years. This
view was recalled on my recent publications of \cite{PaSo} and \cite{So}, and in particular
on stumbling upon \cite{BaFrLe}. 



\begin{thebibliography}{2}
	\bibitem{AD}
		\textsc{Aczél, J.} and \textsc{Daróczy, Z.} (1975). \textit{On measures of information
		and their characterizations.}
		Academic Press, Inc.
	\bibitem{BaFrLe}
		\textsc{Baez, J. C.}, \textsc{Fritz, T.}, and \textsc{Leinster, T.}
		A characterization of entropy in terms of information loss.
		\textit{(preprint, {\tt arXiv:1106.1791})}
	\bibitem{CoTh}
		\textsc{Cover, T. M.} and \textsc{Thomas, J. A.} (1991).
		\textit{Elements of Information Theory}.
		John Wiley \&  Sons, Inc.
	\bibitem{Cs}
		\textsc{Csiszár, I.} (2008).
		Axiomatic Characterizations of Information Measures.
		\textit{Entropy} \textbf{10} 261--273.
	\bibitem{Eb}
		\textsc{Ebanks, B. R.} (2007).
		General solution of the 2-cocycle functional equation on solvable groups
		\textit{Aequationes Math.} \textbf{73} 260--279.		
	\bibitem{Er}
		\textsc{Erdős, P.} (1959).
		A remark on the paper 'On some functional equations' by S. Kurepa
		\textit{Glas. Mat.-Fiz. Astron. Ser. II} \textbf {14} 3--5.
	\bibitem{Fa}
		\textsc{Fadeev, D. K.} (1956).
		Towards the notion of finite entropy of a system of probabilities. (in russian)
		\textit{Uspiehi. matem. nauk.} \textbf{11} 227--231.
	\bibitem{Ha}
		\textsc{Hardy, G.} (1949).
		\textit{Divergent Series},
		Clarendon Press
	\bibitem{Re}
		\textsc{Rényi, A.} (1970).
		\textit{Probability Theory},
		Akadémiai Kiadó
	\bibitem{PaSo}
		\textsc{Paszkiewicz, A., Sobieszek, T.}
		\textit{Additive entropies of partitions, (preprint, {\tt arXiv:1202.4591})}.
	\bibitem{Sh}
		\textsc{Shannon, C. E.}, and \textsc{Weaver, W} (1964).
		\textit{The mathematical theory of communication}.
		The University of Illinois press.
	\bibitem{So}
		\textsc{Sobieszek, T.}
		\textit{Noncontinuous additive entropies of partitions, (preprint, {\tt arXiv:1202.4590})}.
\end{thebibliography}
\end{document}